# Heat Bath Algorithmic Cooling with Spins: Review and Prospects

Daniel K. Park[1, 2], Nayeli A. Rodriguez-Briones[1, 2], Guanru Feng[1, 2], Robabeh R. Darabad[1, 2], Jonathan Baugh[1, 2, 3] and Raymond Laflamme[1, 2, 4, 5]

[1]*Institute for Quantum Computing,* [2]*Department of Physics and Astronomy, and* [3]*Department of Chemistry, University of Waterloo, Waterloo, Ontario, Canada N2L 3G1*
[4]*Perimeter Institute for Theoretical Physics, Waterloo, Ontario, Canada N2J 2W9*
[5]*Canadian Institute for Advanced Research, Toronto, Ontario, Canada M5G 1Z8*

## 1. Introduction

Quantum Information Processing (QIP) has the potential to perform exponentially faster computation and revolutionize current technology by harnessing the systems governed by the laws of quantum mechanics. To reach this goal we need to be able to control imperfections and imprecision occurring when theoretical ideas are implemented in the physical world. Quantum error correction is a theory that aims to bridge this. The first steps towards developing experimental quantum error correction have been taken but there is still more to be done. Application of multiple rounds of Quantum Error Correction (QEC) is an essential milestone towards the construction of scalable QIP devices, but experimental realizations of it are still in their infancy. The requirements for multiple round QEC are high control fidelity and the ability to extract entropy from ancilla qubits. Nuclear Magnetic Resonance (NMR) based quantum devices have demonstrated high control fidelity with up to 12 qubits. Hence, these devices are excellent test beds that can be explored in the lab today for the ideas of quantum control and QEC. More details on NMR QIP can be found in [1,2]. The major challenge in the NMR QEC experiment is to efficiently supply ancilla qubits in highly pure states at the beginning of each round of QEC. This challenging requirement was recently alleviated by Criger et al. in [3]. They showed that QEC could still suppress the error rate using mixed ancilla qubits as long as the polarization or purity of the ancilla qubits is above certain threshold value. Purifying qubits in NMR can be obtained through Heat Bath Algorithmic Cooling (HBAC). It is an efficient method for extracting entropy from qubits that interact with a heat bath, allowing cooling below the bath temperature. In a nutshell, HBAC recurrently applies two steps: Given $n$ number of system qubits each with polarization $\epsilon_0$, cool $n-m$ qubits by compressing entropy into $m$ qubits. The polarization of $m$ qubits is exchanged with the heat bath polarization. By repeating these steps, $n-m$ qubits can attain a final polarization $\epsilon_f$ that is greater than $\epsilon_b$. There is an asymptotic limit for $\epsilon_f$ that depends on $n$ and $\epsilon_b$. Proof-of-principle experiments for 3-qubit HBAC have been performed in a solid state NMR system, demonstrating sufficient level of control to execute multiple rounds of HBAC. However, under typical experimental conditions, nuclear spin polarization at thermal equilibrium is very small and therefore precise control over tens of nuclear spin qubits is required for reaching polarization of order unity on one qubit. For practical HBAC and QEC, coupled electron-nuclear spin systems are more promising than conventional

NMR Quantum Computing (QC), since electron spin polarization is about $10^3$ times greater than that of a proton under the same experimental conditions. This is due to the higher gyromagnetic ratio of the electron spin. Another consequence is that the electron spin relaxation rate is typically about $10^3$ faster than that of nuclear spins. Faster spin relaxation and higher polarization makes the electron an excellent heat bath for cooling nuclear spins.

In this review, we provide an overview on both theoretical and experimental aspects of HBAC focusing on spin and magnetic resonance based systems. The paper is organized as follows. The challenge of preparing nearly pure ancilla qubits in conventional NMR system is discussed in Section 2. Section 3 discusses the theory of HBAC in detail. Section 4 reviews solid state NMR experiments that demonstrated sufficient control fidelity for realizing HBAC, and motivates the use of electron spins by explaining the shortcomings of NMR QC. Section 5 introduces electron-nuclear spin ensemble QC and the prospects of implementing HBAC in this type of system.

## 2. State preparation challenge in ensemble quantum computation

NMR QIP is one example of ensemble quantum computation models, where an ensemble of identical quantum systems is manipulated in parallel and the only measurable quantities are expectation values of certain observables. That is, there is no access to projective measurement. In this section, we review concepts related to spin polarization and present the challenge of preparing nearly pure spin qubits.

### 2.1 Polarization

For a spin at temperature $T$, the occupancy of a state with energy $E$ is calculated by the Gibbs distribution $n(E) = e^{-E/k_B T}/Z$, where $k_B$ is the Boltzmann constant and $Z$ is the partition function. The polarization $\epsilon$ is defined as the population difference between two energy levels normalized by the total number of spins. When the Zeeman energy dominates the energy splitting, the polarization of a spin-1/2 system can be expressed as

$$\epsilon = n(E_0) - n(E_1) = \tanh\left(\frac{\Delta E}{2k_B T}\right) \approx \tanh\left(\frac{\hbar \gamma B_0}{2k_B T}\right), \tag{1}$$

where $\Delta E$ is the energy difference between the two levels, $\gamma$ is the gyromagnetic ratio, and $B_0$ is the strength of the external magnetic field. The equation (1) establishes the relationship between polarization and gyromagnetic ratio, magnetic field strength, and temperature.

### 2.2 Pseudo-pure state

The density matrix describing a spin ensemble at thermal equilibrium can be written in the eigenbasis of $\sigma_z$, corresponding to the direction of the applied static magnetic field, as

$$\rho = \frac{1}{2}\begin{bmatrix} 1+\epsilon & 0 \\ 0 & 1-\epsilon \end{bmatrix} = \frac{1}{2}(\mathbb{I} + \epsilon\sigma_z), \tag{2}$$

where $\mathbb{I}$ is the unit matrix and $\sigma_z$ is the Pauli operator. For $n$ spin qubits, the thermal equilibrium state can be transformed to a pseudo-pure state through non-unitary processes using standard NMR techniques of temporal or spatial averaging [4,5]:

$$\rho_{pps}^n = (1-\alpha)\mathbb{I}_n + \alpha|\psi\rangle\langle\psi|, \tag{3}$$

$$\alpha = \frac{(1+\epsilon)^n - 1}{2^n - 1}, \tag{4}$$

where $|\psi\rangle\langle\psi|$ is a pure state, $\mathbb{I}_n$ is $2^n \times 2^n$ normalized unit matrix, and $\alpha$ quantifies the purity of the state. A typical NMR experiment operates at $B_0 \approx 7\,T$ and room temperature in which nuclear spin polarizations are extremely small ($\epsilon \approx 10^{-5}$ for proton). Moreover, in the absence of methods like algorithmic cooling which can compress entropy, the purity of a pseudo-pure state must decrease exponentially in the number of qubits.

## 2.3 NMR QEC with mixed ancilla qubits

In [3], Criger et al. showed that even if ancilla qubits are not pure, QEC can still suppress the error rate as long as the polarization of ancilla qubits exceeds a certain threshold value. For example, in the conventional 3-qubit QEC code for phase flip error, one can imagine that two ancilla qubits in the NMR experiment are in mixed states with polarizations $\epsilon_1$ and $\epsilon_2$, respectively. The probability amplitude of the lowest energy state of the two qubit has to be greater than 0.5 in order for QEC to suppress the error rate and improve the fidelity of a state exposed to the noisy channel, i.e. $(1+\epsilon_1)(1+\epsilon_2)/4 > 0.5$. If $\epsilon_1 = \epsilon_2 = \epsilon$, then $\epsilon > \sqrt{2} - 1 \approx 0.41$ must be satisfied. This is far above what can be achieved in a reasonable NMR setup. One can imagine having a solid state NMR setup in which the experiment can be carried out at low temperature. However, in order to meet the polarization requirement given above, the temperature must be below $17\,mK$ for $^1$H at a field of $7\,T$. As the temperature is lowered, the nuclear $T_1$ relaxation time is increased and therefore the wait time for thermal state initialization can become impractically long.

In the following sections, we present the main ideas and experimental realizations of heat bath algorithmic cooling, which is a promising tool for preparing nearly pure spin qubits in experiments that are feasible with today's technology.

## 3. Theory of Algorithmic Cooling

Purification of quantum states is essential for realizing fault tolerant quantum information processors. The procedure is needed not only for initializing the physical system for many algorithms, but also to dynamically supply fresh pure ancilla qubits for error correction. For quantum computation models that rely on ensemble of identical systems such as NMR or Electron Spin Resonance (ESR) [6], acquisition of nearly pure quantum states in a scalable manner is extremely challenging. A potential solution is algorithmic cooling (AC), a protocol which purifies qubits by removing entropy from subset of them, while increasing the entropy of the rest [7,8]. An explicit way to implement this idea in quantum computations was given by Schulman et al. [9]. They showed that it is possible to reach polarization of order unity using only a number of qubits which is polynomial in the initial polarization. However, their method was limited by the Shannon bound, which imposes a constraint on the entropy compression step in closed systems.

This idea was further improved by adding contact with a heat bath to pump entropy out of the system and transfer it into the heat bath [10], a process known as Heat Bath Algorithmic Cooling (HBAC). Based on this new idea, many practical cooling algorithms have been designed [11–15]. In short, HBAC purifies qubits by applying alternating rounds of entropy compression and pumping out entropy from the system of interest to a thermal bath. Theoretical details are explained below.

As an example we consider a system consisting of a string of qubits: one target qubit (a spin-1/2) which is to be cooled, one qudit (a spin-$l$) which aids in the entropy compression, and $m$ reset qubits that can be brought into thermal contact with the heat bath. When $l$ is a half integer, having spin-$l$ is equivalent to having $n' = \log_2 d$ qubits where $d$ is the dimension of the Hilbert space of spin-$l$. The spin-1/2 and the spin-$l$ are referred to as computational qubits (Figure 1).

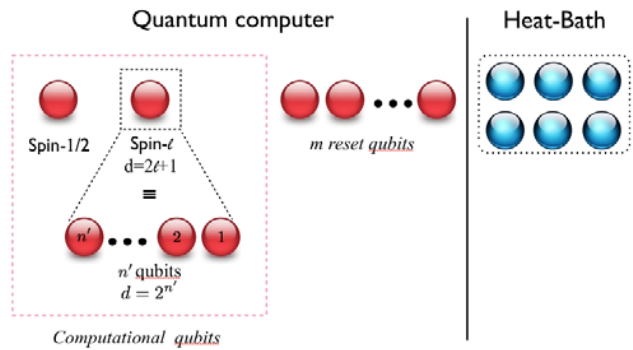

**Figure 1: Heat bath algorithmic cooling can cool the target spin-1/2 by compressing entropy into $d$ - dimensional spin-$l$ (equivalent to a string of $n' = \log_2 d$ qubits if $l$ is a half integer), and exchanging the entropy of spin-$l$ with that of the reset qubits that are in contact with a cold heat bath. The target spin-1/2 and the spin-$l$ are referred to as the computational qubits.**

The idea of the first step of HBAC is to re-distribute the entropy among the string of qubits by applying an entropy compression operation U. This is a unitary (reversible) process that extracts entropy from the computational qubits as much as possible and moves it to $m$ reset qubits, resulting in the cooling of the computational qubits while warming the reset qubits (Figure 2).

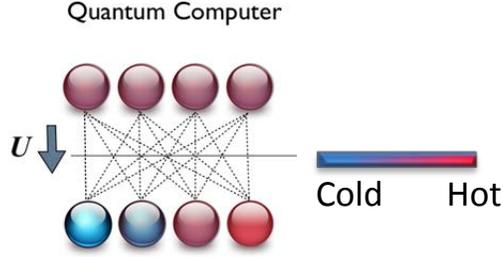

**Figure 2: Entropy compression step. A compression operation U raises the entropy on one side of the system, while lowering it on the other. In the figure, the top part represents the string of qubits before the compression. Dotted lines indicate re-distribution of entropy among all qubits, resulting in the separation of cold and hot regions as shown in the bottom part.**

In the second step, $m$ reset qubits are brought into thermal contact with a heat-bath and reset to the cold bath temperature, resulting in the cooling of the total $n + m$ qubit system. This step is equivalent to tracing over the reset qubits, and replacing them with qubits from the heat bath. The heat bath is assumed to be very large so that the action of qubit-bath interaction on the bath is negligible.

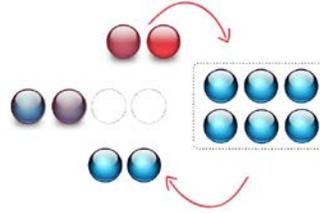

**Figure 3: The refresh step. The reset qubits are brought into thermal contact with a heat bath and the entropy of the qubit system is reduced. In the figure, two reset qubits are used as an example.**

The total effect of these two steps on a system with initial state $\rho$ can be expressed as follows:

$$\rho \xrightarrow{C} \rho' = U\rho U^\dagger \tag{5}$$

$$\rho' \xrightarrow{R} \rho'' = Tr_m(\rho') \otimes \rho_{\epsilon_b}^{\otimes m} \tag{6}$$

where $C$ and $R$ stand for compression and reset, $Tr_m()$ is the partial trace operation over $m$ reset qubits, and $\rho_{\epsilon_b} = \frac{1}{2}\begin{pmatrix} 1 + \epsilon_b & 0 \\ 0 & 1 - \epsilon_b \end{pmatrix}$ represents a qubit possessing the heat bath polarization $\epsilon_b$.

These reversible compression and refreshing steps are iteratively applied until the target qubit reaches the desired temperature, or the cooling limit is reached. The physical requirements for reset and computational qubits are different. A reset qubit should strongly interact with the bath in order to rapidly relax and attain the bath temperature, and a computational qubit should have long relaxation time to remain polarized after being cooled through entropy compression.

In [12], Schulman et al. introduced the optimal (optimal in terms of entropy extraction per cooling step) algorithm for HBAC, the Partner Pairing Algorithm (PPA). The PPA is explained in detail in the following section.

### 3.1 Partner Pairing Algorithm (PPA)

Consider a system with $n - 1$ computational qubits and one reset qubit. Let $\rho$ and $\rho_{\epsilon_b}$ be the density matrices of the $n$ qubits and of the reset qubit after contact with the thermal bath, respectively. $\rho$ can be partitioned into $2^{n-1} \times 2^{n-1}$ blocks $M_{ij}$ by the basis states $\{|0\rangle, |1\rangle\}$ of the reset qubit:

$$\rho = \begin{pmatrix} \begin{bmatrix} \rho_{11} & \rho_{12} \\ \rho_{21} & \rho_{22} \end{bmatrix} & \begin{bmatrix} \rho_{13} & \rho_{14} \\ \rho_{23} & \rho_{24} \end{bmatrix} & \cdots \\ \begin{bmatrix} \rho_{31} & \rho_{32} \\ \rho_{41} & \rho_{42} \end{bmatrix} & \begin{bmatrix} \rho_{33} & \rho_{34} \\ \rho_{43} & \rho_{44} \end{bmatrix} & \\ \vdots & & \ddots & \vdots \\ & \cdots & & \begin{bmatrix} \rho_{2^n-1,2^n-1} & \rho_{2^n-1,2^n} \\ \rho_{2^n,2^n-1} & \rho_{2^n,2^n} \end{bmatrix} \end{pmatrix}$$

(7)

$$= \begin{pmatrix} M_{11} & M_{12} & \cdots \\ M_{21} & M_{22} & \\ & \vdots & \ddots & \vdots \\ & \cdots & & M_{2^{n-1},2^{n-1}} \end{pmatrix},$$

where $M_{ij}$ is the $ij$-block of $\rho$ (for example $M_{11} = \begin{bmatrix} \rho_{11} & \rho_{12} \\ \rho_{21} & \rho_{22} \end{bmatrix}$, $M_{12} = \begin{bmatrix} \rho_{13} & \rho_{14} \\ \rho_{23} & \rho_{24} \end{bmatrix}$, and etc.)

Refreshing the reset qubit (equation (6)) has the effect of changing every block $M_{ij}$ to $M'_{ij}$ as follows:

$$M'_{ij} = \frac{Tr(M_{ij})}{2} \begin{pmatrix} 1 + \epsilon_b & 0 \\ 0 & 1 - \epsilon_b \end{pmatrix}.$$

(8)

In the PPA, entropy compression operation permutes the diagonal elements of the density matrix of the system, rearranging them such that states in increasing lexicographic order have non-increasing probability. For example, the probability amplitude of states starting with 0 (0…00, 0…01, etc.) is increased while that of states starting with 1 is decreased. This operation aims to increase the polarization of the first qubit. The compression can no longer improve the polarization of the first qubit once the states are already ordered as described above.

### 3.2 Illustrative example: PPA for three qubits

We take a system of three qubits (one reset qubit and two computational qubits) as an example to illustrate the PPA. $\rho_0$ is the initial density matrix of the three qubit system and without loss of generality, $\rho_0$ is in a maximally mixed state and the heat bath has polarization $\epsilon_b$.

First, contact between the reset qubit and the heat bath is established. Then the compression operator U permutes the probabilities of the basis states and sorts them in non-increasing order:

$$d(\rho_0) = \frac{1}{8}\begin{bmatrix}1\\1\\1\\1\\1\\1\\1\\1\end{bmatrix} \xrightarrow{R} d(\rho'_0) = \frac{1}{8}\begin{bmatrix}1+\epsilon_b\\1-\epsilon_b\\1+\epsilon_b\\1-\epsilon_b\\1+\epsilon_b\\1-\epsilon_b\\1+\epsilon_b\\1-\epsilon_b\end{bmatrix} \xrightarrow{C} d(\rho''_0) = \frac{1}{8}\begin{bmatrix}1+\epsilon_b\\1+\epsilon_b\\1+\epsilon_b\\1+\epsilon_b\\1-\epsilon_b\\1-\epsilon_b\\1-\epsilon_b\\1-\epsilon_b\end{bmatrix}, \qquad (9)$$

where $d(\rho)$ are the eigenvalues of $\rho$, and $R$ and $C$ stand for refresh and compression steps, respectively. This compression is equivalent to swapping the first computational qubit with the reset qubit. After this iteration, the polarization of the first qubit is increased from 0 to $\epsilon_b$. Upon repeating above two steps again, we effectively swap the second computational qubit with the reset qubit that is at thermal equilibrium with the bath:

$$d(\rho_1) := d(\rho''_0) \xrightarrow{R} d(\rho'_1) = \frac{1}{8}\begin{bmatrix}(1+\epsilon_b)^2\\1-\epsilon_b^2\\(1+\epsilon_b)^2\\1-\epsilon_b^2\\1-\epsilon_b^2\\(1-\epsilon_b)^2\\1-\epsilon_b^2\\(1-\epsilon_b)^2\end{bmatrix} \xrightarrow{C} d(\rho''_1) = \frac{1}{8}\begin{bmatrix}(1+\epsilon_b)^2\\(1+\epsilon_b)^2\\1-\epsilon_b^2\\1-\epsilon_b^2\\1-\epsilon_b^2\\1-\epsilon_b^2\\(1-\epsilon_b)^2\\(1-\epsilon_b)^2\end{bmatrix}. \qquad (10)$$

In this iteration, the polarization of the first computational qubit remains the same, while the polarization of the second qubit is increased from 0 to $\epsilon_b$. If the refresh and compression steps are repeated once more,

$$d(\rho_2) := d(\rho''_1) \xrightarrow{R} d(\rho'_2) = \frac{1}{8}\begin{bmatrix}(1+\epsilon_b)^3\\(1+\epsilon_b)^2(1-\epsilon_b)\\(1+\epsilon_b)^2(1-\epsilon_b)\\(1+\epsilon_b)(1-\epsilon_b)^2\\(1+\epsilon_b)^2(1-\epsilon_b)\\(1+\epsilon_b)(1-\epsilon_b)^2\\(1+\epsilon_b)(1-\epsilon_b)^2\\(1-\epsilon_b)^3\end{bmatrix} \xrightarrow{C} d(\rho''_2) = \frac{1}{8}\begin{bmatrix}(1+\epsilon_b)^3\\(1+\epsilon_b)^2(1-\epsilon_b)\\(1+\epsilon_b)^2(1-\epsilon_b)\\(1+\epsilon_b)^2(1-\epsilon_b)\\(1+\epsilon_b)(1-\epsilon_b)^2\\(1+\epsilon_b)(1-\epsilon_b)^2\\(1+\epsilon_b)(1-\epsilon_b)^2\\(1-\epsilon_b)^3\end{bmatrix}. \qquad (11)$$

In this iteration, the polarization of the first qubit increases to $1.5\epsilon_b - 0.5\epsilon_b^3$.

The gate representations of three entropy compression steps are shown in Figures 4.a, 4.b, and 4.c, respectively. Three iterations complete the first round of 3-qubit HBAC. In the next round, the required compression gates are alternating applications of the operations shown in Figures 4.b and 4.c.

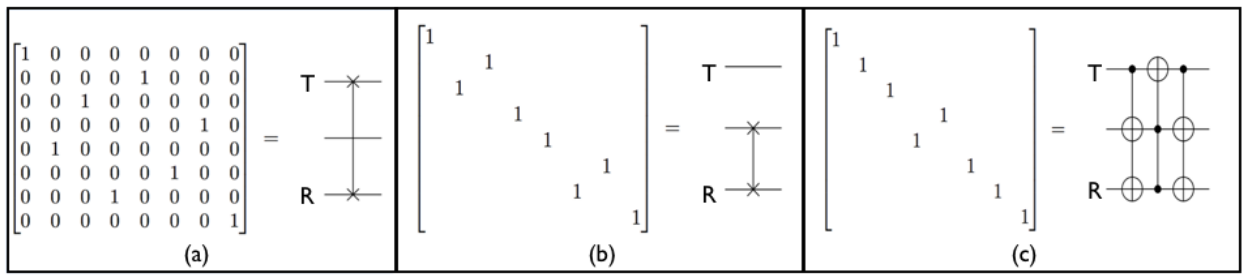

**Figure 4. Matrices and circuit symbols representing the unitary operations of the PPA on three qubits that are initially in a completely mixed state. In the circuit diagram, the top qubit is the target qubit (denoted T) and the bottom qubit is the reset qubit (denoted R). A swap operation is represented as two X's located on the qubits that are exchanged connected by a vertical line. A controlled-not gate is denoted by a dot and an open circle connected by a vertical line. The open circle is on the target qubit of the controlled-not operation, and the dot is on the controlled qubit. (a) In the first iteration, the compression gate swaps the target qubit and the reset qubit. (b) In the second iteration, the second qubit and the reset qubit are swapped. (c) The third iteration boosts the first qubit polarization to $1.5\epsilon_b - 0.5\epsilon_b^3$. From the second round of HBAC and on, entropy compressions are the repetition of the second and third gates of the first round.**

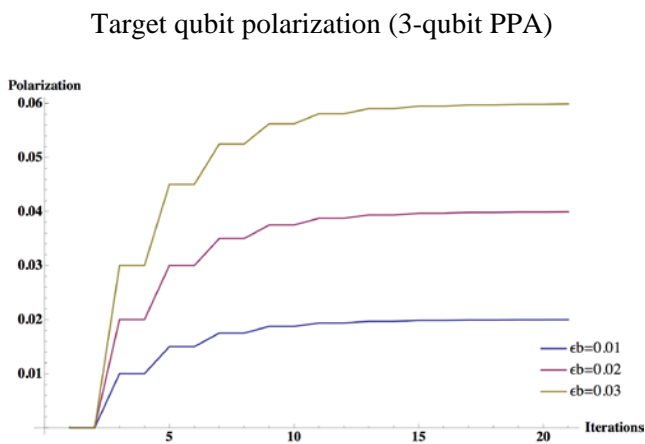

Target qubit polarization (3-qubit PPA)

**Figure 5: Evolution of the target qubit polarization under the PPA method, using a system of 3 qubits, for three values of heat bath polarization $\epsilon_b$. Each iteration consists of a reset and a compression procedure. Note the asymptotic polarization is $2\epsilon_b$, as expected for $\epsilon_b \ll 1/2$ in the case of three qubits.**

The evolution of the polarization of the first qubit under the PPA with $\epsilon_b \ll 1/2$ is shown in

Figure 5. The circuit asymptotically boosts the polarization on the first qubit up to twice the heat bath polarization; this limit is discussed in the next section.

An interesting question is to know what is the maximum achievable cooling using this method, and how many iterations of HBAC are necessary to obtain a certain polarization.

## 3.3 Cooling limit

Through numerical simulations, Moussa [16] (see also [12]) observed that if $\epsilon_b \ll 1/2^{n-2}$, where $n$ is the number of computational plus reset qubits, the maximum polarization the target qubit can have is $2^{n-2}\epsilon_b$. But when $\epsilon_b > 1/2^{n-2}$, a polarization close to one can be reached. Recently, the cooling limit of the PPA (starting with completely mixed qubits) was solved analytically: the maximum polarization of the target qubit can be expressed as a function of the number of computational and reset qubits and the heat bath polarization [17,18]. This exact solution is consistent with the upper bound found by Schulman et al. [14]. The cooling limit corresponds to the stage at which it is not possible to continue extracting entropy from the system, i.e. when the state of the system is not changed by the compression and refresh steps. The system achieves this limit asymptotically, converging to a steady state. Starting from the completely mixed state, the density matrix always remains in a diagonal form after each HBAC step. In general, the state of the computational qubits $\rho_{comp}$ is therefore completely described by its diagonal elements: $d(\rho_{comp}) = (A_1^t, A_2^t, \ldots, A_{2d}^t)$ after $t$ iterations of HBAC.

The cooling limit is reached when there is no operation that can compress the entropy of the computational qubits, or equivalently, when the diagonal elements of the total state are already sorted. This limit occurs when the elements of $\rho_{comp}$ satisfy the following condition:

$$A_i(1-\epsilon_b)^m = A_{i+1}(1+\epsilon_b)^m. \tag{12}$$

This condition together with normalization gives the exact solution of the steady state of the computational qubits, $\tilde{\rho}_{comp}$:

$$d(\tilde{\rho}_{comp}) = A_1(1, Q, Q^2, Q^3, \ldots, Q^{2d-1}), \tag{13}$$

where $A_1 = \frac{1-Q}{Q(1-Q^{2d})}$ and $Q = \left(\frac{1-\epsilon_b}{1+\epsilon_b}\right)^m$.

The maximum achievable polarization corresponds to the polarization of the state in the cooling limit. From the steady state, the maximum polarization of the target qubit is as follows:

$$\epsilon_{max} = \frac{(1+\epsilon_b)^{md} - (1-\epsilon_b)^{md}}{(1+\epsilon_b)^{md} + (1-\epsilon_b)^{md}}. \tag{14}$$

In the limit of low heat bath polarization, $\epsilon_b \ll 1/(md)$, the polarization of the steady state is proportional to $md$, in agreement with simulations. As the value of the heat bath polarization increases beyond $md$, the final polarization grows arbitrarily close to 1. The final polarization of the target qubit as a function of the heat bath polarization is shown in Figure 6 for different numbers of qubits.

In order to see how quickly $\epsilon_{max}$ approaches 1, we introduce $\delta_{max} = 1 - \epsilon_{max}$. Using equation (14), $\delta_{max}$ can be expressed as:

$$\delta_{max} = \frac{2}{e^{md \ln\left(\frac{1+\epsilon_b}{1-\epsilon_b}\right)} + 1} = \frac{2}{e^{m2^{n'} \ln\left(\frac{1+\epsilon_b}{1-\epsilon_b}\right)} + 1}. \tag{15}$$

This expression shows that the maximum polarization reaches 1 exponentially in the size of the Hilbert space $d$ (or doubly exponentially in $n'$, the number of computational qubits excluding the target qubit).

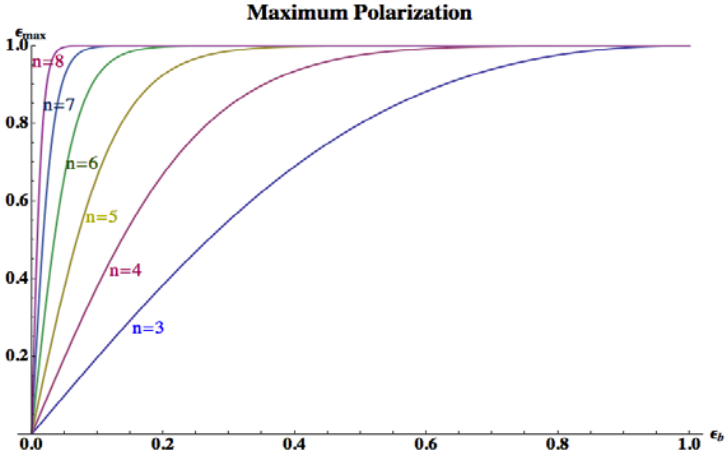

**Figure 6:** Maximum polarization achievable for the target qubit versus heat bath polarization $\epsilon_b$ and the number of qubits. The maximum polarization increases doubly exponentially in $n'$, the number of computational qubits excluding the target qubit. This plot shows the results for $n = 3, 4, 5, 6, 7,$ and $8$, where $n$ is the sum of computational qubits and one reset qubit (i.e. $n' = 1, 2, 3, 4, 5,$ and $6$).

## 4. Experimental Algorithmic Cooling with NMR QIP

Experimental realization of algorithmic cooling requires high fidelity control and the ability to reset qubits. Liquid State NMR (LSNMR) QIP has successfully demonstrated precise quantum control up to 12 qubits, and hence it can provide sufficient quantum control for the experimental demonstration of algorithmic cooling. Nevertheless, the only way to reset qubits in LSNMR relies on spin-lattice relaxation, characterized by the time scale $T_1$. Reset qubits must have very short $T_1$ values compared to the computational qubits. One can imagine a molecule in which one nuclear spin has a very rapid $T_1$ and the reset to 99% of the thermal polarization can be achieved by waiting $5T_1$. However, other spin qubits must have much slower $T_1$ process in order to maintain their polarizations during the reset step. Furthermore, the short $T_1$ on the reset qubit limits its $T_2$ and the fidelity of control. Despite these limitations, some preliminary steps towards full PPA have been experimentally realized in LSNMR by using protons ($^1$H nuclei) as reset qubits and $^{13}$C nuclei as computational qubits [19,20]. These experiments showed selective reset operations to polarize all three spin qubits close to the bath temperature. Nevertheless, the final compression step which polarizes a target qubit colder than the heat bath was not implemented. Meanwhile, Chang et al. implemented cooling solely by the final compression gate on three fluorines in $C_2F_3BR$ using LSNMR. Full implementation of HBAC in LSNMR was accomplished much later in [21].

On the other hand, a network of dipolar coupled spins in Solid State NMR (SSNMR) offers a reset step that does not require a relaxation process in the system of interest. A large number of dipolar coupled spins can be thought of as a spin bath, and other spins can be brought in thermal contact with the bath and reach thermal equilibrium at the spin bath temperature. Moreover, SSNMR experiments can be operated at low temperature, providing a higher bath polarization. In this section, we review the experimental demonstration of 3-qubit algorithmic cooling using a molecular single crystal.

## 4.1 Brief review of Solid-State NMR QC

Solid state NMR QIP makes use of the techniques developed in LSNMR QIP, and offers several advantages: the decoherence rates can be made slow using refocusing techniques, while spin-spin couplings much larger than in LSNMR can be exploited to realize faster quantum gates [22].

Features of the internal Hamiltonian of SSNMR that differ from LSNMR are the anisotropic chemical shift and dipole-dipole couplings between nuclei. The anisotropic chemical shift should be described by a tensor $\boldsymbol{\delta}$. In the secular approximation (at large dc magnetic field), the form of the dipole-dipole interaction Hamiltonian depends on whether the interacting spins belong to the same isotopic species or not, and can be written as follows:

$$\text{Homonuclear: } H_D^{ij} = d_{ij}(3\hat{I}_z^i \hat{I}_z^j - \hat{\boldsymbol{I}}^i \cdot \hat{\boldsymbol{I}}^j), \tag{16}$$

$$\text{Heteronuclear: } H_D^{ij} = d_{ij}(2\hat{I}_z^i \hat{I}_z^j), \tag{17}$$

$$d_{ij} = -\hbar \frac{\mu_0}{4\pi} \frac{\gamma_i \gamma_j}{r_{ij}^3} \frac{3\cos^2\theta_{ij} - 1}{2}, \tag{18}$$

where $\mu_0$ is the permeability of free space, $\gamma_i$ is the gyromagnetic ratio of spin $i$, $r_{ij}$ is the distance between interacting spins, and $\theta_{ij}$ is the angle between the vector connecting the two spins and the external magnetic field. There are also J-couplings in SSNMR, which are usually an order of magnitude smaller than the dipole-dipole couplings, and for which the isotropic component cannot be distinguished from dipolar couplings. Nuclei with $S > 1/2$ interact with external electric field gradients, a phenomenon known as quadrupolar interaction. In this chapter, we limit our discussions to spin-1/2 systems, and hence quadrupolar interactions do not appear in the internal Hamiltonian.

Because of the orientation dependence of the internal Hamiltonian and the ensemble nature of NMR QIP, the solid sample should be a single crystal where each unit cell contains only one molecule (or >1 magnetically equivalent molecules).

## 4.2 SSNMR algorithmic cooling experiment

Experimental HBAC using SSNMR was first demonstrated by Baugh et al. in 2005 [23]. They implemented the PPA for three qubits using a single crystal of malonic acid $CH_2(COOH)_2$ (Figure 7) as the quantum processor at $B_0 = 7.1\ T$ and room temperature.

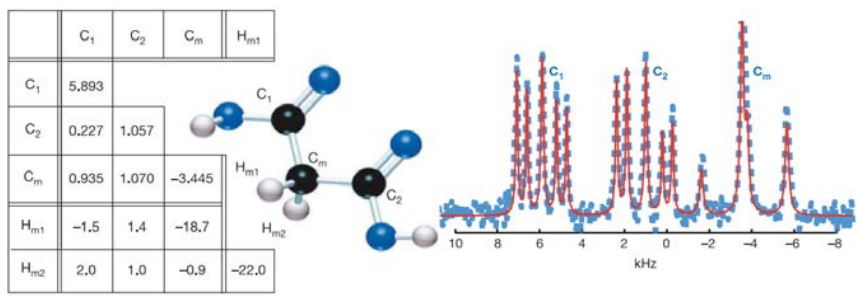

Figure 7: Characteristics of the dilute 3-$^{13}$C malonic acid spin system. On the right is the $^1$H-decoupled $^{13}$C spectrum. The blue dashed line is the experimental NMR spectrum, and the red line is the fit. The peaks are grouped into three multiplets which can be assigned to $C_1$, $C_2$ and $C_3$. In each mutiplet, the central peak comes from the natural abundance of $^{13}$C, and is inconsequential for QIP purposes. The table gives the parameters of the Hamiltonian. The diagonal values are chemical shifts and the off-diagonal values are dipolar couplings. All values are in kHz. Reprinted by permission from Macmillan Publisher Ltd: Nature 438, 470-473 (2005), copyright (2005).

In each unit cell of malonic acid, there are two molecules that are related by a centre of symmetry and magnetically equivalent to each other (P$\bar{1}$ space group). The molecules in which all three carbons are isotopically labelled as $^{13}$C (3-$^{13}$C) were used as quantum information processors, while the 100% abundant $^1$H spins in the crystal were used as the heat bath. The concentration of 3-$^{13}$C molecules in the crystal was 3.2%. There were also about 1.1% molecules with one $^{13}$C spin and about $(1.1\%)^2$ molecules with two $^{13}$C spins due to natural abundance of $^{13}$C spins, the latter being a small enough concentration to neglect. The small fraction of molecules with one $^{13}$C produces detectable NMR signal, but these are inconsequential for QIP purposes. The structure of the molecule, spin Hamiltonian parameters used for the experiment that were obtained from spectral fitting, and $^1$H-decoupled $^{13}$C spectrum are shown in Figure 7.

The quantum circuit for 3-qubit PPA is shown in Figure 8. The experiment was designed to increase the polarization of $C_1$, and $C_m$ was the qubit interacting with the heat bath. As discussed in Section 3, the algorithm combines two steps: refreshing step and reversible polarization compression step. The refreshing step, illustrated in Figure 8, is realized via the SWAP gate between $C_m$ and $H_{m1}$. As shown in the spin Hamiltonian parameters in Figure 7, the orientation of the crystal with respect to the magnetic field was chosen in such a way that only $H_{m1}$ had a large coupling with $C_m$. The other heteronuclear couplings were negligibly small and all the homonuclear couplings were refocused during the refresh step.

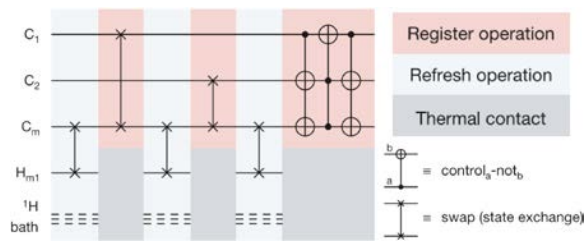

**Figure 8: The schematic circuit of the solid-state NMR PPA experiment. The circuit contains three iterations: three refresh steps and three reversible polarization compression steps, which are labelled as 'register operation'. Reprinted by permission from Macmillan Publisher Ltd: Nature 438, 470-473 (2005), copyright (2005).**

The SWAP is implemented by 'time-suspension' sequence [24] (Figure 9) which induces effective spin-exchange Hamiltonian in the form

$$H_{eff}^{ij} = \frac{d_{ij}}{3}(\hat{\mathbf{I}}^i \cdot \hat{\mathbf{I}}^j), \qquad (19)$$

where spin $i$ and spin $j$ correspond to $C_m$ and $H_{m1}$. The evolution of duration $\tau = 3/(2d_{ij})$ under this spin-exchange Hamiltonian is equivalent to the SWAP.

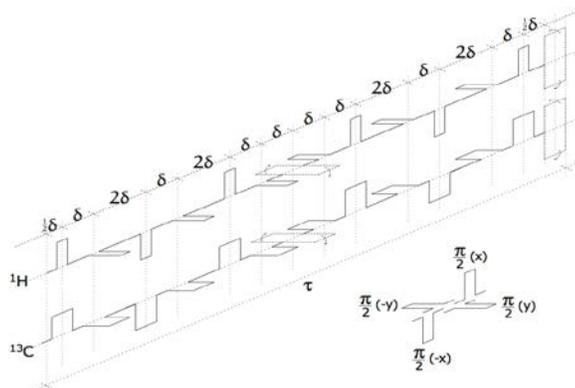

**Figure 9: The 'time-suspension' sequence of hard pulses for implementing the $^1$H-$^{13}$C SWAP gate. Each rectangle corresponds to a $\pi/2$ rotation about a Bloch sphere axis ($\pm x$ or $\pm y$) denoted by the orientation of the rectangle, and $\delta$ is the shortest time delay between pulses. The sequence decouples homonuclear couplings, and transforms the heteronuclear Hamiltonian to an exchange Hamiltonian. The total time of the pulse sequence is $\tau = 3/(2d_{ij})$. This figure is reproduced from [16].**

Due to experimental imperfections, only about 83% of the thermal $^1$H polarization is transferred to $C_m$. The final compression step was realized using strongly modulating pulses [25]. The pulse was designed to be robust to radio frequency (RF) field inhomogeneity. During the application of permutation gates, $^1$H spins in the crystal were strongly decoupled from $^{13}$C spins, and 'spin-locked' by a transverse, phase-matched RF field, which preserved the $^1$H polarization and allowed the $H_{m1}$ to re-equilibrate with the hydrogen bath through spin diffusion mediated by hydrogen-hydrogen dipolar couplings. In the beginning of the experiment, all $^{13}$C spins were initialized in completely mixed states by rotating the thermal polarization to a transverse Bloch sphere axis and dephasing it. After first five steps of the experiment, ideally the polarization of all three $^{13}$C spins should equal to the bath polarization $\epsilon_B$. Then in principle, the final gate boosts the polarization of $C_1$ to $1.5\epsilon_B$ to first order in $\epsilon_B$. In the experiment, the final polarization of $C_1$ was $1.22 \pm 0.03$ times the bath polarization. Major sources of error were a non-ideal process of spin diffusion that prevented perfect refresh of $C_m$ as shown in Figure 10, and RF control imperfections. In spite of experimental errors, $C_1$ attained polarization higher than that of the heat bath.

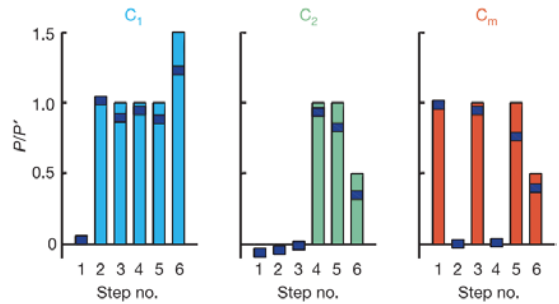

Figure 10: Theoretical and experimental qubit polarizations at each step for $C_1$, $C_2$, and $C_3$. Bars indicate ideal qubit polarizations; shaded bands are experimental values. The thickness of shaded bands indicates experimental uncertainty. **Reprinted by permission from Macmillan Publisher Ltd: Nature 438, 470-473 (2005), copyright (2005).**

In 2008, Ryan et al. experimentally demonstrated four rounds of algorithmic cooling that consist of nine iterations (Figure 11) in the same experimental system (malonic acid) [26]. Perfect control without decoherence would theoretically result in improving one of the $^{13}$C polarizations to $1.94\epsilon_B$ after four rounds of the PPA.

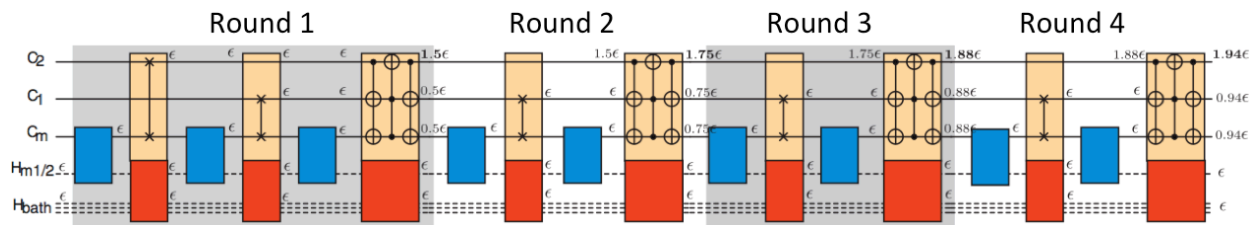

Figure 11: The schematic circuit of the four-round PPA experiment. Theoretical values of attainable polarizations are provided with respect to the heat bath polarization $\epsilon$ at each stage of the experiment. **Reprinted figure with permission from C. A. Ryan, O. Moussa, J. Baugh, and R. Laflamme, Phys. Rev. Lett. 100, 140501 (2008). Copyright (2008) by the American Physical Society.**

The experiment resembles the work in [23] in the sense that three $^{13}$C spins were used as the computational qubits and the abundant $^1$H spins were used as the heat bath. But the refresh and the polarization compression steps were implemented differently. In the refresh step, the polarization was transferred from $H_{m1}$ and $H_{m2}$ at the same time. The orientation of the sample was chosen in such a way that both $H_{m1}$ and $H_{m2}$ have large dipolar couplings with $C_m$. Instead of applying a multi-pulse sequence to realize the refreshing step, cross polarization (CP) [27] which was found to be better in preserving the heat bath polarization was used. The permutation gates applied on $^{13}C$'s were numerically optimized using the GRadient Ascent Pulse Engineering (GRAPE) algorithm [28]. The pulse design takes several error sources into considerations: distributions of the static magnetic field and the RF control field, and the finite bandwidth of the NMR probe resonant circuit. Also, the optimized pulses were corrected for nonlinearities in the pulse generation and transmission to the sample through a procedure that measures the RF pulse at the sample and corrects it via feedback. With all these tools, the level of control was improved and enabled four rounds of PPA. The polarization of each $^{13}C$ at the end of each round is shown in Figure 12.

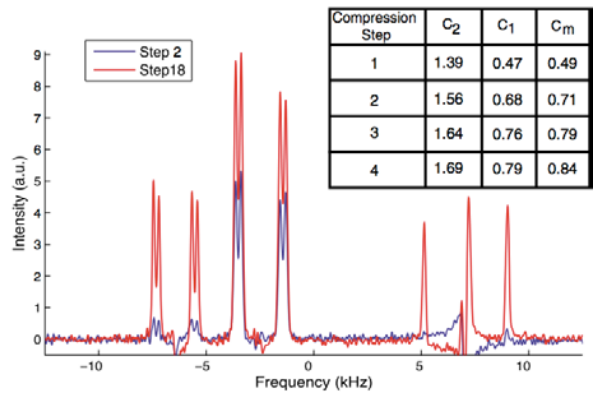

**Figure 12: Table of the measured polarization (with respect to the initial refresh step) of each spin after each round of PPA. The spectra show a comparison of the first refresh step (swapped to $C_2$) and the final signal after four rounds of PPA. It shows clearly that the polarization of $C_2$ is boosted well beyond the bath polarization. Reprinted figure with permission from C. A. Ryan, O. Moussa, J. Baugh, and R. Laflamme, Phys. Rev. Lett. 100, 140501 (2008). Copyright (2008) by the American Physical Society.**

After four rounds of PPA, the polarization of $C_2$ reached $1.69\epsilon_B'$, where $\epsilon_B'$ is the polarization of $C_m$ after the first reset. Experimental error was dominated by two factors—imperfection of $^1H$ decoupling and the network of dipolar coupled protons in the bath leading to a non-ideal process of spin diffusion.

These experiments were significant milestones towards implementation of active error correction in solid state spin ensemble QIP. It demonstrated sufficient control fidelity to realize HBAC to prepare an ancilla qubit whose polarization is higher than the cold bath polarization. Now the control tools are available, and what remains a challenge of experimental QEC is to identify a system that provides a heat bath that can be polarized to much higher values than the thermal nuclear bath.

### 4.3 Limitations to HBAC Using NMR

The SSNMR experiment discussed in this section successfully improved polarization of one nuclear spin beyond the heat bath polarization. Nevertheless, the ultimate goal of dynamical supply of nearly pure ancilla qubits for QEC is still far from reach. For example, the 3-qubit dephasing QEC code requires two ancilla qubits with polarizations of 0.41 at minimum (see Section 2.3). In order to achieve this polarization on one nuclear spin at room temperature and $B_0 \approx 7\,T$, perfect quantum control on 17 coupled $^1$H with no losses due to decoherence would be required. The condition can be relaxed if the heat bath temperature is colder. However, operating NMR HBAC at cryogenic temperatures is not an ideal solution for preparing cold heat bath since $T_1$ of nuclear spins become undesirably long at low temperature. This naturally leads to exploiting electron spin since the electron gyromagnetic ratio is much higher than that of nuclei which results in higher thermal equilibrium polarization and faster $T_1$ relaxation.

In the following section, we discuss implementation of HBAC using a combination of electron and nuclear spin resonance in hyperfine-coupled quantum processors.

## 5. HBAC with Hyperfine-Coupled Electron-Nuclear Spin Ensemble

The fundamentals of ESR QC are analogous to NMR QC, and many of the techniques used for manipulating nuclear spins can also be applied to control electrons. One obvious advantage is that higher gyromagnetic ratio of an electron $\gamma_e$ (about 660 times greater than that of proton) leads to higher polarization. Decoherence and relaxation rates also scale with $\gamma$ and hence electron $T_1$ relaxation rate is about 3 orders of magnitude larger than that of nuclei. Thus the electron spin is an excellent candidate for the reset qubit, and the reset can be done simply by waiting for a time about $5T_1$. Anisotropic hyperfine interaction is an advantage for designing nuclear quantum gates since it provides a control handle for fast manipulations of nuclear spins. However, in the case of HBAC, strong anisotropic hyperfine interaction can be a disadvantage because electron $T_1$ relaxation process induces nuclear polarization decay in the presence of anisotropic hyperfine interaction. If the interaction is strong, the loss of nuclear polarization while resetting the electron can be significant. Fortunately, one can choose the crystal orientation to reduce the anisotropic hyperfine coupling strength so that the nuclear spin decay induced by electron $T_1$ is small enough to allow cooling of a target spin species below bath temperature. We discuss the spin Hamiltonian in more detail in the following section, and also discuss the crystal orientation selection for realistic implementations.

### 5.1 The Electron-Nuclear Spin Hamiltonian

The spin Hamiltonian for a 1 electron, k nuclear spin-1/2 system can be written as

$$H = \beta_e g_{\mu\nu} B_\mu \hat{S}_\nu + \sum_{n=1}^{k} \left( A^n_{\mu\nu} \hat{S}_\mu \hat{I}^n_\nu - \gamma_n \hat{I}^n_\mu B_\mu \right), \qquad (20)$$

where $\hat{S}$ and $\hat{I}$ represent electron and nuclear spin operators, $\beta_e$ is Bohr magneton, $\vec{B}$ is the external magnetic field and $\gamma_n$ is the gyromagnetic ratio for nuclear spin $n$. The second rank tensors $g$ and $A^n$ are the electron g-tensor and the $n^{th}$ nuclear spin hyperfine coupling tensor, respectively. The nuclear dipole-dipole interaction is neglected since it is typically at least two orders of magnitude weaker than the hyperfine terms.

Pulsed ESR spectrometers are classified according to the frequency of the microwave source. Most commonly, ESR experiments are conducted at X-band (8-12 GHz) frequency, mainly due to the relatively low cost of microwave amplifiers and other components in this frequency range. In X-band ESR, the electron Zeeman interaction is the dominating term of the Hamiltonian. By convention, the coordinate system is chosen such that $\vec{B} = B_0 \hat{z}$, and the electron spin is quantized along that direction. When the magnitudes of nuclear Zeeman energy and the hyperfine interaction are comparable and much smaller than the electron Zeeman energy, the spin Hamiltonian is well approximated as

$$H = \omega_S \hat{S}_z + \sum_{n=1}^{k} \left[ -\omega_I^n \hat{I}_z^n + \hat{S}_z \left( a_n \hat{I}_z^n + b_n \hat{I}_x^n \right) \right]. \qquad (21)$$

Here, $\omega_S = \beta_e g_{zz} B_0$ and $\omega_I^n = \gamma_n B$ are electron and nuclear Larmor frequencies respectively, and $a_n = A^n_{zz}$ and $b_n = \sqrt{(A^n_{zx})^2 + (A^n_{zy})^2}$.

There are two schemes for achieving universal control in electron nuclear systems. In the first scheme, the nuclear spins are directly manipulated by external RF pulses that are on resonance with NMR transition frequencies. This technique is known as Electron Nuclear Double Resonance (ENDOR) [29]. The second approach is to exploit the anisotropic hyperfine coupling and indirectly manipulate nuclear spins via microwave (MW) pulses acting on the electron. For brevity, we will name the latter approach Anisotropic Hyperfine Control (AHC). In the following section, we explain how to achieve universal control of electron-nuclear coupled systems through AHC in more detail.

## 5.2 Indirect Control via Anisotropic Hyperfine Coupling

The anisotropy of hyperfine coupling permits nuclear spin manipulation solely by irradiating MW pulses at electron spin transitions. The control universality of a 1 electron, N nuclear spin coupled system via anisotropic hyperfine interaction was proved in [30], and demonstrated experimentally in [31] for a single nuclear spin qubit gate and in [32] for a gate involving two nuclear spin qubits. The advantage of the indirect control technique is that it simplifies the instrumentation as additional RF excitations are not needed, and faster gate implementation relative to ENDOR can be achieved when the hyperfine coupling strength exceeds the Larmor frequency of the nucleus in a given external field. Here we use 1 electron, 1 nuclear spin system as an example to illustrate the idea. In the presence of $\vec{B} = B_0 \hat{z}$ and the hyperfine interaction, the nuclear spin is quantized along the direction of an effective field

$$\vec{B}_n = \left(B_0 \pm \frac{a}{2\gamma_n}\right)\hat{z} \pm \frac{b}{2\gamma_n}\hat{x}, \tag{22}$$

and the $\pm$ sign depends on whether the electron spin is parallel (spin up) or antiparallel (spin down) to the external field. As a consequence, when $b \neq 0$, the direction of the nuclear spin quantization axis is dictated by the electron spin state. We introduce $\theta_\uparrow = \arctan\left(\frac{-b}{a+2\omega_I}\right)$ and $\theta_\downarrow = \arctan\left(\frac{-b}{a-2\omega_I}\right)$ to denote the angle of nuclear spin quantization axes from $\hat{z}$ axis, and $\Theta = (\theta_\uparrow - \theta_\downarrow)/2$. Then the eigenstates of the coupled spin system are [29]:

$$
\begin{aligned}
|1\rangle &= |\uparrow\rangle \otimes \left(\cos\left(\frac{\theta_\uparrow}{2}\right)|\uparrow\rangle - \sin\left(\frac{\theta_\uparrow}{2}\right)|\downarrow\rangle\right), \\
|2\rangle &= |\uparrow\rangle \otimes \left(\sin\left(\frac{\theta_\uparrow}{2}\right)|\uparrow\rangle + \cos\left(\frac{\theta_\uparrow}{2}\right)|\downarrow\rangle\right), \\
|3\rangle &= |\downarrow\rangle \otimes \left(\cos\left(\frac{\theta_\downarrow}{2}\right)|\uparrow\rangle - \sin\left(\frac{\theta_\downarrow}{2}\right)|\downarrow\rangle\right), \\
|4\rangle &= |\downarrow\rangle \otimes \left(\sin\left(\frac{\theta_\downarrow}{2}\right)|\uparrow\rangle + \cos\left(\frac{\theta_\downarrow}{2}\right)|\downarrow\rangle\right).
\end{aligned}
\tag{23}
$$

In the eigenbasis of the spin Hamiltonian shown in equation (21), the rotating-frame electron control Hamiltonian $H_c = \omega_1 \hat{S}_x$ becomes

$$\widetilde{H}_c = \frac{\omega_1}{2}\begin{bmatrix} 0 & 0 & \cos(\Theta) & -\sin(\Theta) \\ 0 & 0 & \sin(\Theta) & \cos(\Theta) \\ \cos(\Theta) & \sin(\Theta) & 0 & 0 \\ -\sin(\Theta) & \cos(\Theta) & 0 & 0 \end{bmatrix}. \tag{24}$$

From equation (24), one can see that the control Hamiltonian is able to induce all transitions between any eigenstates of the electron spin up manifold and the electron spin down manifold provided $\Theta \neq n\pi/2$ where n is an integer and eigenstates are non-degenerate. The energy level connectivity can be represented as a graph, and the complete connectivity of the graph generated by the control Hamiltonian and non-degenerate energy levels guarantee universal control of the system [31,33,34]. Since the spin Hamiltonian does not consider nuclear-nuclear dipolar interactions, the idea presented for 1 electron, 1 nuclear spin system can be easily extended to larger number of nuclear spins, provided that suitable and distinct hyperfine couplings exist. The loss of nuclear spin polarization due to electron $T_1$ in the presence of anisotropic hyperfine interaction can be intuitively understood from equation (24). When $\sin(\Theta)$ is non-zero, there is a finite probability for the nuclear spin to flip to its low energy state through electron-nuclear double spin relaxation.

## 5.3 HBAC Simulations in Electron-Nuclear Coupled Systems

The ancilla qubits for the 3-qubit dephasing QEC code must be polarized to at least 41% in order to correct errors at all. In theory, X-band ESR algorithmic cooling at $4.2K$ can yield greater than 40% polarization on one nuclear spin using 1 electron and 4 hyperfine coupled nuclei. This is a significant reduction in the number of necessary qubits compared to NMR case at room temperature, which required 17 protons (both examples here assume error-free controls).

In this section, we explain how to experimentally implement the control, and present proof-of-principle simulation results that reflect more realistic control and decoherence parameters to examine the feasibility of HBAC in the electron-nuclear coupled systems. Simulations were carried out for both ENDOR and AHC control schemes. We consider gamma-irradiated malonic acid in which one carbon is isotopically labelled as $^{13}C$ (Figure 13) as an example. The idea can, in principle, be extended to larger electron-nuclear spin ensemble systems.

In the simulation, electron $T_1$ and $T_2$ processes are modelled as a Markovian dynamical map, and simulated by solving a master equation of the Lindblad form [35,36]. Inhomogeneous line broadening of the electron spin resonances is taken into account by averaging the simulation over a set of spin Hamiltonians in which the magnitude of electron Zeeman energy is a Lorentzian-distributed random variable. We use experimentally measured electron $T_1$ and $T_2$ values to determine the Lindblad operators, and the measured ESR line width to determine the $T_2^*$ Hamiltonian distribution.

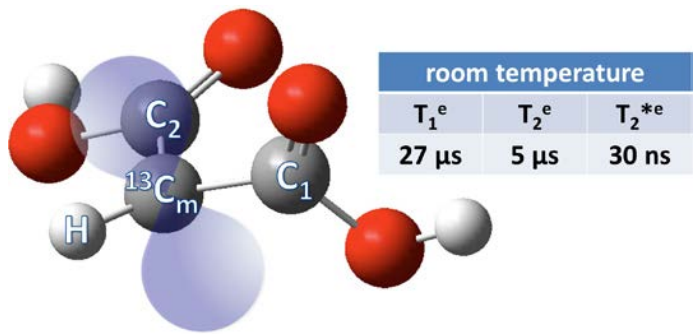

**Figure 13: Schematic of the gamma-irradiated malonic acid with three qubits (electron, $\alpha$-$^1H$ and methylene $^{13}C$). The electron density distribution is represented by the blue shaded region. Relaxation parameters of the electron $T_1^e$, $T_2^e$ and $T_2^{*e}$ are measured in an X-band pulsed ESR spectrometer at room temperature.**

### 5.3.1 HBAC using ENDOR Control

In ENDOR, nuclear spin flip transitions are directly excited by RF pulses oscillating at the nuclear frequencies. As shown in Figure 14, all the gates used in the PPA can be decomposed into controlled-not (CNOT) gates that are realized by transition selective $\pi$-pulses. For example, a CNOT gate with the electron as the control qubit and a nuclear spin as the target is implemented by irradiating the sample with RF pulse at the frequency that corresponds to the energy difference between $|\uparrow_e \uparrow_n\rangle$ and $|\uparrow_e \downarrow_n\rangle$, at pulse amplitude $\omega$ for duration $\tau$ such that $\omega\tau = \pi$. A Toffoli gate that flips the electron if two nuclei are both in the spin up state can be realized by exciting the transition between $|\uparrow_e \uparrow_n \uparrow_n\rangle$ and $|\downarrow_e \uparrow_n \uparrow_n\rangle$ with a MW pulse of amplitude $\omega$ for duration $\tau$ such that $\omega\tau = \pi$.

Since the spin Hamiltonian depends on the dc magnetic field orientation with respect to the crystallographic axes, we aim to select an orientation that maximizes the polarization improvement on the target spin qubit. In the ENDOR experiment, the orientation selection requirements are as follows: (1) the electron-nuclear double spin flip rate is as slow as possible by minimizing $\sin(\theta)$ in equation (24); (2) all allowed transitions (as opposed to forbidden transitions) should be separated by more than the relevant ESR or NMR line width; and (3) the bandwidth of microwave control is narrow enough to obtain control faster than $T_2^e$, but wide enough to irradiate all relevant ESR transitions.

Using $g$, $A^H$, and $A^C$ that are determined from continuous-wave ESR (CW ESR) measurement at X-band, we determined a crystalline orientation in which all conditions above are reasonably well satisfied. We performed a simulation of 3-qubit ENDOR algorithmic cooling designed to increase the polarization of the electron (see Figure 14).

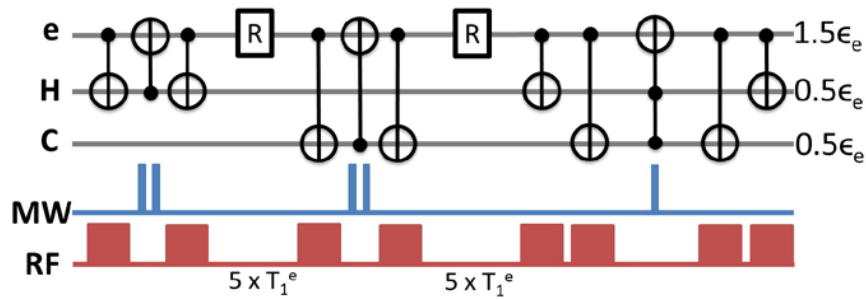

**Figure 14: Quantum circuit and corresponding pulse sequence for the 3-qubit PPA using pulsed ENDOR to control the electron, $\alpha$-proton and methylene $^{13}$C of gamma irradiated malonic acid. Blue and red rectangles indicate MW and RF pulses that selectively excite particular transitions for a single electron spin flip and a single nuclear spin flip, respectively. The reset (indicated by R) is accomplished by waiting for 5 times the $T_1$ of the electron. This single round of HBAC boosts the polarization of the electron spin approximately 1.5 times compared to its thermal equilibrium polarization.**

In principle, a single round of 3-qubit HBAC increases the target spin polarization to 1.5 times the heat bath polarization. The simulation takes finite pulse width, $T_1^e$, $T_2^e$ and $T_2^{*e}$ into account. The reset is done by waiting for 5 times the $T_1$ of the electron. The simulation uses $50\,ns$

Gaussian shaped pulses for MW, and 15 $\mu s$ and 60 $\mu s$ square pulses for RF. Gaussian shaped pulses are used in the MW channel in order to maintain selectivity of a particular transition while exciting over the full ESR line width. The amplitudes of the RF pulses are chosen to reflect typical RF amplifier output power levels. Since the RF pulses are similar in duration to $T_1^e$ at room temperature, the algorithm does not yield polarization increase at room temperature. A solution is to increase $T_1^e$ by performing the experiment at low temperature. We use the experimentally determined value of $T_1^e = 2.6\ ms$ at $43K$. Taking into account all the relaxation parameters and simulating the experiment at $T = 43K$ yield a final polarization (after one round of the PPA) of 1.36 times the bath polarization.

### 5.3.2 HBAC Using Anisotropic Hyperfine Control

When the universal control is achieved through anisotropic hyperfine coupling and electron spin excitation [31,32], the orientation selection criteria are modified as following: (1) the electron-nuclear double spin flip rate (i.e. forbidden rate) is strong enough that nuclear gates can be implemented quickly compared to the electron $T_2$; (2) however this forbidden transition rate must also be weak enough that it does not significantly speed up the nuclear $T_1$ process; and (3) the frequencies of all transitions (allowed and forbidden) should be separated from each other by at least the ESR line width to achieve high fidelity control.

Using the same electron g-tensor and hyperfine interaction tensors, a crystal orientation can be found that satisfies new conditions above. The GRAPE algorithm [28] is then used to design the swap and compression gates via microwave control of the electron spin. The corresponding quantum circuit is illustrated in the figure below. All three pulses are designed to have 99% unitary fidelity, and the pulse lengths are $840\ ns$, $840\ ns$, and $900\ ns$ for electron-$^1$H swap, electron-$^{13}$C swap, and compression, respectively.

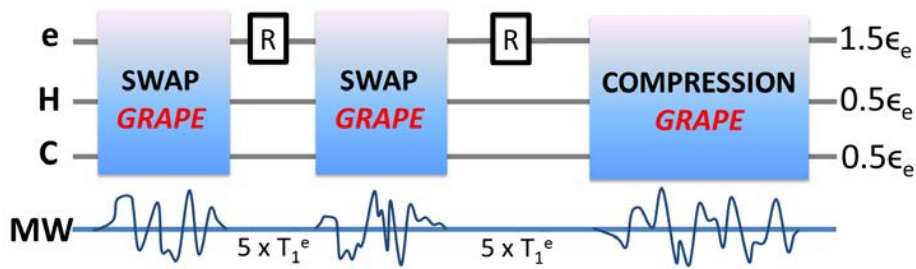

**Figure 15: Quantum circuit and corresponding pulse sequence for the 3-qubit PPA using electron, $^1$H and $^{13}$C spins and microwave-only control. Swap and compression gates are realized by shaped pulses found using the GRAPE algorithm** [28]**. The reset steps are done by waiting for 5 times the $T_1$ of the electron. One round of HBAC boosts the electron spin polarization to 1.5 times its thermal equilibrium polarization.**

The simulated final polarization of the electron after one round of the PPA is 1.21 times the bath polarization. The polarization improvement here is worse compared to the previous ENDOR simulation results due to the fact that $T_2^e$ is only about 5 times longer than the duration of each GRAPE pulse, and that nuclear polarizations decay faster due to anisotropic coupling during the reset steps. On the other hand, one immediate advantage of this control scheme compared to ENDOR experiment is that the experiment can be done at room temperature since pulse durations are much shorter than room temperature $T_1^e$. Moreover, the ability to implement nuclear gates solely through MW pulses greatly simplifies the experimental hardware.

## 5.4 Prospects: Exploiting Larger Hilbert Spaces and the High Field Regime

Achieving high control fidelity and experimental demonstration of electron-nuclear spin HBAC in the proof-of-principle level remains to be experimentally demonstrated. Nevertheless, given sufficient control, HBAC can potentially be explored using molecules with a greater number of nuclear spin qubits coupled to an electron. One example is the diphenyl nitroxide radical (see Figure 16) [37,38]. Diphenyl nitroxide as an open-shell molecular sample is an extremely stable nitroxide radical with electron spin-1/2. Mixed single crystals of diphenyl nitroxide in benzophenone are grown with a sufficiently dilute concentration of the radical to suppress electron-electron dipolar interactions. Diphenyl nitroxide isostructurally replaces the benzophenone molecules. This provides 10 protons that are hyperfine coupled to the electron and are spectrally distinguishable through ENDOR. Nitrogen can be isotopically labelled as $^{15}N$ to provide an additional strongly coupled nuclear spin. This spin system therefore affords the possibility of implementing HBAC with up to 12 qubits [39].

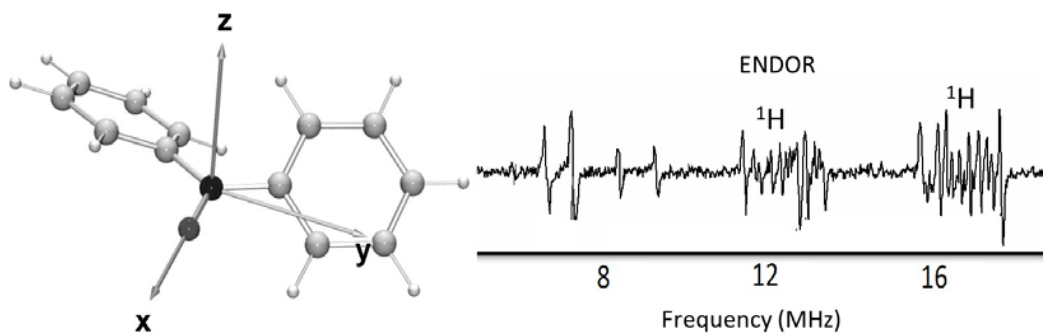

**Figure 16: Left: Molecular structure of the diphenyl nitroxide radical. Crystal axes are labelled as x, y, and z. Right: ENDOR spectrum of diphenyl nitroxide in a crystalline benzophenone matrix. 20 ENDOR peaks in the neighborhood of 12 MHz and 16 MHz correspond to the nuclear frequencies of 10 proton spins. This spectrum indicates that all 10 nuclear spins are spectrally distinct and may be selectively controlled using ENDOR. These figures are reproduced from** [39]

The maximum polarization that can be reached by one nuclear spin in 12-qubit HBAC at X-band is 0.67 at room temperature. It can be improved to more than 0.99 at liquid nitrogen temperature $77K$, and a polarization arbitrarily close to 1 is possible at liquid helium temperature $4.2K$. Figure 17 shows the theoretical polarization of a target spin as a function of the number of HBAC iterations at various temperatures. One iteration of HBAC consists of a reset and a compression. Note that in order to achieve high polarization, a very large number of iterations is required unless the experiment is performed at a temperature below $77K$ (see Figure 17).

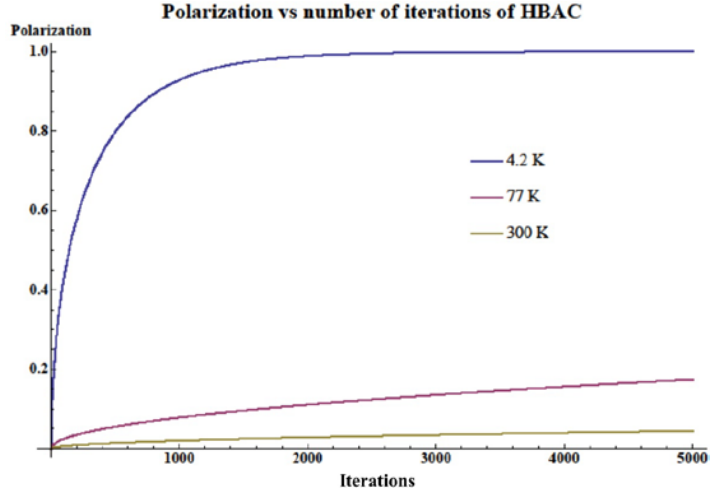

**Figure 17: Polarization of a target nuclear spin in the diphenyl nitroxide radical plotted as a function of the number of HBAC iterations at three different temperatures. Each iteration consists of one refresh and one compression step.**

HBAC in high field ESR, such as W-band in which the Larmor frequency of the free electron is about 94 GHZ, is also a possible solution to reach necessary polarization for QEC. Although high fidelity unitary quantum control in the high frequency regime has not been demonstrated, in theory only four rounds of 3-qubit HBAC at $8K$ can supply a qubit with 50% polarization, and above 80 % and 97% polarizations at $4K$ and $2K$, respectively.

## 6 Conclusions and Prospects

NMR experiments on spin ensembles have provided an excellent ground for developing and testing ideas of quantum information processing in the few-qubit regime, owing to the ability to implement quantum control. However, it has lacked the ability to prepare high purity ancilla qubits that are essential for QEC. The challenge for systems like NMR in which projective measurement is not available is that qubit initialization is normally only attainable via thermal equilibration, which results in very low polarizations (i.e. highly mixed qubit states). Dynamic nuclear polarization [40,41] and algorithmic cooling are two means by which non-equilibrium polarizations can be achieved, with this review having discussed the latter. In this chapter, we reviewed Heat Bath Algorithmic Cooling (HBAC), an efficient method for extracting entropy from spin qubits, allowing cooling below the cold bath temperature. The theory of HBAC has been extensively studied, and sufficient quantum control to operate several rounds of HBAC in a 3-qubit system was demonstrated using solid state NMR. In standard NMR QIP, achieving qubit polarizations necessary for QEC is a practical impossibility, given the small values of thermal equilibrium nuclear polarizations even at cryogenic temperatures. Electron-nuclear hyperfine coupled systems are more promising for HBAC than conventional NMR processors since the electron spin has about 3 orders of magnitude larger Zeeman energy, and thus its polarization and spin-lattice relaxation rates are correspondingly higher. The fast $T_1$ process is exploited for the reset step in HBAC. Experiments to demonstrate HBAC in electron-nuclear systems are in progress, but in this review we have presented realistic simulation results indicating that polarization enhancements on the order of the theoretical values are possible.

HBAC is an implementation independent approach and it can also be applied in other spin systems and other QIP implementations. For example, nitrogen-vacancy centers in diamond or a photo-excited triplet state can provide highly polarized spins at room temperature by optical pumping and dynamic nuclear polarization [42–44], and HBAC can be utilized to further purify spin qubits in these systems.

**Acknowledgements**

This work is supported by CIFAR, Industry Canada, and NSERC. We thank Dr. Tal Mor and Dr. Yossi Weinstein for helpful discussions, and Dr. Rolf Horn for proofreading the manuscript. NRB acknowledges CONACYT-COZCyT and SEP for support.